\documentclass[12pt]{article}

\usepackage{amsmath, amsfonts, amssymb, bm}
\usepackage{graphicx}
\usepackage[a4paper, total={6.25in, 8.45in}]{geometry}
\usepackage[colorlinks=true, citecolor=blue, linkcolor = black]{hyperref}

\usepackage[utf8]{inputenc}
\usepackage[T1]{fontenc} 
\usepackage{authblk}
\usepackage{multicol}
\usepackage{helvet}

\usepackage{natbib}
\setcitestyle{authoryear}

\newcommand{\odevar}{\bm{y}}
\newcommand{\odeparam}{\bm{\psi}}
\newcommand{\param}{\bm{\theta}}
\newcommand{\dat}{\mathcal{D}}
\newcommand{\method}{M}
\newcommand{\approxquantity}{\mathcal{Y}}
\newcommand{\tol}{\epsilon}
\newcommand{\atol}{\tol_{\text{abs}}}
\newcommand{\rtol}{\tol_{\text{rel}}}
\newcommand{\normedError}{v}

\newcommand{\E}[2]{\mathbb{E}_{#1}\left[ #2 \right] }
\newcommand{\deriv}[2]{\frac{\text{d} #1}{\text{d} #2}}
\newcommand{\partderiv}[2]{\frac{\partial #1}{\partial #2}}
\newcommand{\titlefont}[1]{{\fontfamily{cmss}\selectfont #1}}
\newcommand{\maxNorm}[1]{\left \lVert #1  \right\rVert_{\text{max}}}

\title{ \titlefont{An importance sampling approach for reliable and efficient inference in Bayesian ordinary differential equation models }}
\author[1]{\normalsize \titlefont{Juho Timonen}}
\author[1]{\normalsize \titlefont{Nikolas Siccha}}
\author[2]{\normalsize \titlefont{Ben Bales}}
\author[1]{\normalsize\titlefont{Harri Lähdesmäki}}
\author[1]{\normalsize \titlefont{Aki Vehtari}}
\affil[1]{\normalsize \titlefont{Department of Computer Science, Aalto University, Finland}}
\affil[2]{\normalsize \titlefont{Earth Institute, University of Columbia, New York, USA}}
\date{\normalsize \titlefont{\today}}

\begin{document} 

\pagestyle{plain}
\maketitle 

\begin{abstract}
Statistical models can involve implicitly defined quantities, such as solutions to nonlinear ordinary differential equations (ODEs), that unavoidably need to be numerically approximated in order to evaluate the model. 
The approximation error inherently biases statistical inference results, but the amount of this bias is generally unknown and often ignored in Bayesian parameter inference. We propose a computationally efficient method for verifying the reliability of posterior inference for such models, when the inference is performed using Markov chain Monte Carlo methods. 
We validate the efficiency and reliability of our workflow in experiments using simulated and real data, and different ODE solvers. We highlight problems that arise with commonly used adaptive ODE solvers, 
and propose robust and effective alternatives which,  accompanied by our workflow, can now be taken into use without losing reliability of the inferences.
\end{abstract}

\section{Introduction}
    
Implicitly defined quantities that depend on unknown parameters introduce challenges when they are involved in statistical models. Examples of such quantities are solutions to 
parameterized algebraic equations, optimization problems, integrals, or ordinary differential equations (ODEs). They generally do not have a closed form given the parameters, and to evaluate the model likelihood they have to be approximated using numerical methods \citep{suli2003}. In the Bayesian context, model inference is commonly done by sampling the posterior distribution of the parameters using Markov chain Monte Carlo (MCMC) \citep{brooks2011} techniques. The theory behind MCMC assumes that the model likelihood can be computed exactly, but this is not the case if numerical approximations are required. Some numerical routines can estimate their own error,
but as the true error is not known and its magnitude varies in different parts of the parameter space, it is difficult to predict how it affects the MCMC posterior draws. As this bias has not had a lot of attention in the literature, unaware users can blindly use default configurations of the numerical methods implemented in software packages. 

Numerically solving an ODE system is a computationally intensive task, and often dominates the cost of one unnormalized posterior density evaluation in Bayesian ODE models. Computational requirements are amplified by the fact that MCMC inference typically requires a large number of these ODE solutions, and this number can be several orders of magnitude larger than the posterior effective sample size (ESS) due to the need to warm-up the sampler and high auto-correlation of intermediate draws. The ratio of ESS to the number of unnormalized posterior density evaluations can be drastically increased by using gradient-based MCMC methods such as Hamiltonian Monte Carlo (HMC) \citep{duane1987}, and using the gradient information is essential in the case of high-dimensional parameter spaces or complex posterior geometries in order to achieve good sampling performance. In modern statistical software such as Stan \citep{StanTeam}, PyMC \citep{salvatier2016}, and Turing.jl \citep{ge2018}, gradients are computed using automatic differentiation \citep{bell2008, baydin2018}. This poses a challenge for ODE models, for which computing the gradient is computationally significantly more demanding and subtle than the plain likelihood evaluation.

Classic numerical integrators for solving ODE systems are iterative methods that discretize the integral, and their accuracy and stability depend on the discretization step size \citep{hairer1993, griffiths2010, suli2003}. In theory, the error can be made arbitrarily small by using a step size that approaches zero, but this is not possible in practice due to limited computational resources and floating point arithmetic. Smaller step size means more evaluations of the ODE right-hand side (RHS) function, which means more computation. Selecting the step size therefore involves balancing between a reasonable computation time and good accuracy of the approximation.

Adaptive integrators remove the burden of selecting the step size, as they can tune it automatically. These methods estimate their own local error, and adapt their step size so that given tolerances are satisfied. However, this does not give any guarantees about validity of the related statistical inference results. Moreover, requiring more accuracy typically causes the solver to adapt to smaller step size values, which leads to more computation. The problem of step size selection has thus been replaced by the problem of selecting the tolerances. Regardless, adaptive solvers are the most commonly used methods in statistical software, and have been included in various probabilistic programming and machine learning frameworks that implement gradient-based MCMC samplers with automatic differentiation.

To our best knowledge, there exist no generally applicable frameworks for validating the reliability of an approximate numerical method that is needed for posterior density evaluations in MCMC inference. For ODE solvers, one approach is to gradually use lower and lower step sizes (or stricter and stricter tolerances for adaptive solvers) during inference, until posterior estimates do not change appreciably anymore. However, repeating MCMC sampling like this quickly becomes computationally very expensive. 
\cite{capistran2016} recognized the model that uses a numerical approximation as a different model than the true model with the exact ODE solution. In the special case of a Gaussian observation model and a certain type of fixed-step solver with step size $h$, they showed that the Bayes factor of the two models approaches one with the same rate as the numerical solution approaches the true solution, as $h \rightarrow 0$. Relying on this, they estimated the marginal likelihood of the true model based on first estimating it for approximate models, with different $h$, and then extrapolating to $h=0$ using linear regression. However, the required several marginal likelihood approximations are difficult to perform in high dimensions and \cite{capistran2016} only demonstrated their method in one-dimensional parameter spaces. 

Probabilistic numerical methods \citep{hennig2015} view the numerical problems probabilistically and can give uncertainty estimates for the solution. \cite{teymur2021} used Gaussian process regression to estimate a distribution for the exact solution given a series of approximations of different accuracy. These methods are designed for performing a fixed numerical problem probabilistically, and it is not clear how to use them in Bayesian inference of models that contain numerical problems with unknown parameters. For the probabilistic parameter inference problem of ODE models, there exist strategies that completely avoid numerically integrating the ODE \citep[see e.g.][and references threrein]{barber2014}. However, such approaches lack convergent numerical methods and thus are not asymptotically approximating the true ODE solutions.

We present an efficient, reliable, and generally applicable strategy for MCMC inference of models which require numerically approximating parameter-dependent quantities. It uses importance weights to correct the biases that result from using numerical approximations. These importance weights are cheap to compute compared to the cost of MCMC sampling, and we can diagnose their success using Pareto smoothed importance sampling \citep{vehtari2021}. The proposed method is straightforward to implement in probabilistic programming languages, as it does not require modifications to standard MCMC algorithms or classic numerical solvers. We demonstrate its benefits in ODE model inference, using both adaptive and non-adaptive ODE solvers.

\section{Methods}

\subsection{Ordinary differential equations}
\label{s: odes}
\subsubsection{Initial value problems}
Ordinary differential equation (ODE) models are routinely used in various fields of science to model dynamic phenomena. A $D$-dimensional ODE system with state variables $\odevar(t) \in \mathbb{R}^D$ is defined as 
\begin{equation}
    \label{eq: ode}
    \deriv{\odevar(t)}{t} = f_{\odeparam}\left(\odevar(t), t\right),
\end{equation}
where the right-hand side function (RHS) $f_{\odeparam}: \mathbb{R}^D \times \mathbb{R} \rightarrow \mathbb{R}^D$ has parameters $\odeparam$. These parameters can be a subset of all parameters of a Bayesian model, in which ODE systems usually appear in the form of initial value problems (IVPs). This means that an initial value $\odevar^{(0)} := \odevar(t_0)$ is explicitly defined (either a known value or a model parameter), and evaluating the likelihood of the data requires solving $\odevar(t)$ at several time points $t > t_0$. The solution is implicitly defined by Eq.~\ref{eq: ode} and the initial value, and can be written using the integral formula
\begin{equation}
    \label{eq: ode_solution}
    \odevar(t) = \odevar^{(0)} + \int_{t_0}^{t} f_{\odeparam}\left(\odevar(t'), t'\right) \text{d}t',
\end{equation}
which according to the Picard-Lindelöf theorem has a unique solution assuming some smoothness conditions\footnote{We only consider problems where these conditions are satisfied.} for $f_{\odeparam}$ \citep{hairer1993}. However, the integral rarely has an explicit closed form and has to be approximated numerically.

\subsubsection{ODE solvers}
A myriad of different methods exist for numerically approximating the integral in Eq.~\ref{eq: ode_solution}. We use $\odevar^{\method}(t)$ to denote the solution given by a numerical method $\method$. A general strategy used by method $\method(h)$ with fixed step size $h > 0$ is to first compute $\odevar^{\method(h)}(\tau_j)$ on a grid $\tau_j = t_0 + jh$, $j \in \{0, 1, 2, \ldots\}$. This can be done by setting $\odevar^{\method(h)}(\tau_0) = \odevar^{(0)}$ and iteratively computing $\odevar^{\method(h)}(\tau_{j+1})$ for $j \geq 0$ using some update rule. After this, some interpolation method can be used if the solution is required at a time point $t$ which is not on the grid \citep{hairer1993}. Numerical methods are generally required to be \textit{convergent}, meaning that the global error $\left\lVert \odevar^{\method(h)}(t) - \odevar(t)\right \rVert$ must approach zero as $h \rightarrow 0$. A method is called convergent of order $p$ if this happens at rate $\mathcal{O}(h^p)$ \citep{hairer1993}.

 Smaller step sizes $h$ will give a more accurate solution, but require more iterations and therefore more computation. In practice, one would like to set the step size small enough to achieve good precision, but large enough to avoid unnecessary computation. Adaptive step size methods try to automatically adapt the step size by estimating their own error. As the global error is difficult to estimate, software implementations are usually based on estimating the local truncation error, i.e. the error induced by a single step \citep{griffiths2010}. The step size is adapted so that user-supplied absolute ($\atol$) and relative ($\rtol$) tolerances in the estimated local truncation error are satisfied. While these methods remove the burden of selecting $h$ from the user, the user must still supply the absolute and relative tolerances. These tolerances have virtually the same trade-off as the step size selection itself; lower tolerances give better accuracy, but require smaller step sizes and therefore more computation. 
 
 ODE solvers are generally either explicit or implicit. For explicit solvers, the next state is computed explicitly based on the current state. Implicit solvers tend to perform significantly better for stiff problems \citep{hairer1996}, but the downside is that they require numerically solving a system of algebraic equations on each step. This has to be done using for example Newton iteration, which has its own stopping criterion that affects the result. The ODE solvers used in our experiments are described in more detail in Appendix~\ref{s: ode_solver_details}.

\subsubsection{Sensitivity analysis}
\label{s: ode_sens}
Gradient-based MCMC requires computing the gradient of the unnormalized posterior density. In modern probabilistic programming frameworks, gradients are computed using automatic differentiation (AD) \citep{baydin2018, margossian2019}. This involves building a differentiable computation graph for the target density, where all operations on inputs are recorded. However, computing the sensitivities efficiently and reliably in AD frameworks is not straightforward when iterative numerical solvers are involved \citep{bell2008, margossian2019, rackauckas2021}.

There are three main ways to integrate ODE solves into the computation graph built for AD. The \emph{direct method} treats the ODE solve similarly as any other sequence of operations and directly records these operations into the computation graph. We use this method for all non-adaptive solvers in our experiments. The other two methods rely on \emph{continuous sensitivity analysis} \citep{rackauckas2021}. This approach is based on the fact that applying the chain rule of differentiation to the ODE system (Eq.~\ref{eq: ode}) gives
\begin{equation}
    \deriv{}{t} \deriv{\odevar(t)}{\odeparam} = \deriv{}{\odeparam} \deriv{\odevar(t)}{t} = \deriv{}{\odeparam} f_{\odeparam}(\odevar(t), t) = \partderiv{f_{\odeparam}(\odevar(t), t)}{\odevar}  \deriv{\odevar(t)}{\odeparam} + \partderiv{f_{\odeparam}(\odevar(t), t)}{\odeparam},
\end{equation}
from which we get an additional ODE system 
\begin{equation}
\label{eq: sensitivity_system}
    \deriv{}{t} S(t) = \partderiv{f_{\odeparam}(\odevar(t), t)}{\odevar}  S(t) + \partderiv{f_{\odeparam}(\odevar(t), t)}{\odeparam} 
\end{equation}
with $DP$ dimensions. 
\emph{Forward} continuous sensitivity analysis solves this extended system simultaneously with the original system using the same adaptive numerical ODE solver. This can be implemented so that also the extended system needs to satisfy the given tolerances, but more sophisticated rules are used with implicit solvers that require also Newton iteration \citep{feehery1997}. In our experiments, we use forward continuous sensitivity analysis for all adaptive solvers.
\emph{Adjoint} continuous sensitivity analysis \citep{margossian2019, rackauckas2021} first solves only the original ODE system forward in time, and uses the obtained solution to solve a different additional system backward in time to get the sensitivities. In this method the additional system has only dimension $D$, so it theoretically scales better with respect to the number of parameters. However, this method is even harder to configure, and we leave it for future work.

\subsection{Bayesian models with numerical approximations}
We consider MCMC inference of models that define a posterior $p(\param \mid \dat) \propto p(\dat \mid \param)p(\param)$, where $p(\dat \mid \param)$ is the likelihood of data $\dat$ given parameters $\param$, and $p(\param)$ is the prior. The goal of inference is commonly the computation of expectations of the form
\begin{equation}
\label{eq: post_expectation}
    \E{p(\param \mid \dat) }{\varphi(\param)} = \int \varphi(\param) p(\param \mid \dat) \text{d} \param,
\end{equation}
which can be for instance model predictions or parameter estimates, determined by the function $\varphi$. When MCMC is used to obtain posterior draws $\param_s$, $s = 1, \ldots, S$, the integral can be estimated as
\begin{equation}
    \E{p(\param \mid \dat)}{\varphi(\param)} \approx \frac{1}{S} \sum_{s=1}^S \varphi(\param_s).
\end{equation}
We focus on models whose unnormalized posterior density depends on $N$ intermediate quantities $\approxquantity_{n}(\param)$, collected in the list $\approxquantity(\param) = \left\{ \approxquantity_{n}(\param) \right\}_{n=1}^N$. We define 
\begin{equation}
  \mathcal{P} \left(\approxquantity(\param), \param \right) := p(\dat \mid \param)p(\param)
\end{equation}
to denote the unnormalized density. Quantities $\approxquantity(\param)$ can be defined implicitly through equations that have no closed form solution or for other reasons have to be numerically approximated as $\approxquantity^{\method}_{n}(\param) \approx \approxquantity_{n}(\param)$ for each $n = 1, \ldots, N$, where $\method$ denotes the approximation method. Since $\approxquantity(\param)$ can only be approximated, $\mathcal{P} \left(\approxquantity(\param), \param \right)$ cannot be evaluated exactly and therefore it is not possible to use MCMC to sample from $p(\param \mid \dat)$ exactly. Instead, MCMC will sample from some distribution $p^\method(\param \mid \dat) \propto \mathcal{P} \left(\approxquantity^{\method}(\param), \param \right)$. It is therefore crucial to develop methods that can correct this bias and inform users if the approximation method $\method$ is so inaccurate that it renders the correction impossible, and re-running MCMC with a more accurate method is needed.
 
Various numerical methods $\method$ have some control parameters that can be used to tune their accuracy. Examples of such methods are ODE solvers $\method = \method(h)$, where $h > 0 $ is the step size. In ODE solvers, the implicitly defined quantities are $\approxquantity_n(\param) =  \odevar_{\odeparam}(t_n)$, where $\odeparam$ are a subset of all parameters $\param$ and $\odevar_{\odeparam}(t_n)$ is the solution to an ODE initial value problem with parameters $\odeparam$, at time $t_n$. The corresponding numerical approximate solution with method $\method$ we denote $\odevar_{\odeparam}^{\method}(t_n) \approx \odevar_{\odeparam}(t_n)$. 

\subsection{Importance sampling approach}

In importance sampling, draws
$\param'_s$, $s = 1, \ldots, S$ are obtained from another distribution $q(\param)$, and the expectation  (Eq.~\ref{eq: post_expectation}) can be estimated as
\begin{equation}
\label{eq: importance_sampling}
    \mathbb{E}_{p(\param \mid \dat) } \left[ \varphi(\param) \right] \approx \frac{\sum_{s=1}^S r_s  \varphi(\param'_s) }{\sum_{s=1}^S r_s },
\end{equation}
where $r_s = \frac{p(\param'_s \mid \dat) }{q(\param'_s) }$ are called importance ratios/weights. Our approach is to use MCMC importance sampling
with $q(\param) = \mathcal{P} \left(\approxquantity^{\method}(\param), \param \right)$, where $\method$ is the approximation method used during MCMC. The importance ratios are
\begin{equation}
    r_s^{\method} = \frac{\mathcal{P} \left(\approxquantity(\param'_s), \param'_s \right)}{\mathcal{P} \left(\approxquantity^{\method}(\param'_s), \param'_s \right)},
\end{equation}
but as we cannot evaluate $\mathcal{P} \left(\approxquantity(\param'_s), \param'_s \right)$, exact importance sampling is not possible. Instead, we use ratios
\begin{equation}
\label{eq: computable_is_ratios}
    r_s^{\method,\method^*} = \frac{\mathcal{P} \left(\approxquantity^{\method^*}(\param'_s), \param'_s \right)}{\mathcal{P} \left(\approxquantity^{\method}(\param'_s), \param'_s \right)},
\end{equation}
where $\method^*$ is a more accurate method than $\method$. We require that $\method^*$ is a convergent numerical method, meaning that for any fixed $\param$,
\begin{equation}
    \approxquantity^{\method^*}_{n}(\param) \rightarrow \approxquantity_{n}(\param)
\end{equation}
for all $n = 1, \ldots, N$, as $\method^*$ is made more and more accurate. This means that
\begin{equation}
    \mathcal{P} \left(\approxquantity^{\method^*}(\param), \param \right) \rightarrow \mathcal{P} \left(\approxquantity(\param), \param \right)
\end{equation}
at each point $\param$. Consequently, the ratios $r_s^{\method,\method^*}$ converge to the exact ratios $r_s^{\method}$, and posterior estimates (Eq.~\ref{eq: importance_sampling}) converge, too. Our approach therefore is to incrementally increase the accuracy of $\method^*$ until the maximum absolute error
\begin{equation}
\label{eq: max_absolute_error}
    \text{MAE}^{\method, \method^*} = \max_{s} \left\{ \max_n \left\{ \maxNorm{ \approxquantity_n^{\method}(\param_s') - \approxquantity_n^{\method^*}(\param_s') } \right\} \right\}
\end{equation}
has converged\footnote{We use $\maxNorm{\bm{x}} = \max_i |x_i|$ to denote the maximum norm for vectors.}. In general high-dimensional parameter spaces of ODE models, this step can be done with negligible computational effort compared to the initial MCMC sampling. Furthermore, this analysis can be conveniently done simultaneously as we assess whether MCMC needs to be run again with a more accurate method $\method$.

\subsection{Pareto smoothing and diagnostics}
\label{s: psis}
Importance sampling requires that the distribution $q(\param)$ is sufficiently similar to the target distribution $p(\param \mid \dat)$, so that the nominator and denominator in Eq.~\ref{eq: importance_sampling} would have finite variance \citep{geweke1989, geweke2005, koopman2009}. Pareto smoothed importance sampling (PSIS) \citep{vehtari2021} modifies the raw ratios so that the estimator of the expectation has finite variance and asymptotic normality in a wider range of problems \citep{vehtari2021}. However, any modifications to the ratios cannot correct for a $q(\param)$ that is too far from $p(\param \mid \dat)$, which in our case means that the method $\method$ used during MCMC sampling has to be sufficiently accurate. Importantly, PSIS provides a diagnostic that we can use to assess whether this is the case.

In PSIS, a generalized Pareto distribution (GPD) is fitted to match the tail of the empirical distribution of the ratios $r_s^{\method,\method^*}$. The probability density function of the GPD is
\begin{equation}
    \label{eq: gpd}
    p_{\text{GPD}}(x \mid u, \sigma, k) =
    \begin{cases}
    \frac{1}{\sigma}\left( 1 + \frac{k(x-u)}{\sigma} \right)^{-\frac{1}{k} - 1}, \ &k \neq 0 \\
    \frac{1}{\sigma}\exp\left(\frac{x-u}{\sigma} \right), \ &k = 0
    \end{cases}
\end{equation}
where $u \in \mathbb{R}$ is a location parameter, $k \in \mathbb{R}$ is a shape parameter and $\sigma > 0$ is a scale parameter. The cutoff value $u = \hat{u}$ is first determined as explained in \cite{vehtari2021}, and $k, \sigma$ are then fitted to the empirical distribution of the tail $r_s^{\method,\method^*} | r_s^{\method,\method^*} > \hat{u}$. The latter part can be done using the method by \cite{zhang2009}, and overall fitting the GPD parameters is computationally very cheap.

The estimated shape parameter $\hat{k}$ can be used as a diagnostic to determine if the importance sampling estimate (Eq.~\ref{eq: importance_sampling}) is reliable \citep{vehtari2021}. Values $\hat{k}<0.7$ indicate small errors with high probability and good convergence rate with increasing sample size \citep{vehtari2021}. Moreover, as we increase the accuracy of $\method^*$, we can study the convergence of $\hat{k}$ as an additional safeguard metric to assess whether the distribution of importance ratios, and therefore any posterior estimates, have converged. 

\subsection{The proposed workflow}
\label{s: proposed_workflow}
The proposed algorithm for MCMC inference of models that require numerical approximations, can be summarized in the following steps.

\begin{enumerate}
  \item Select a reasonable approximation method $\method$.
  \item Sample parameter draws $\param_s'$, $s=1, \dots, S$, using MCMC with $\method$ as the approximation method.
  \item Compute $\text{MAE}^{\method, \method^*}$ (Eq.~\ref{eq: max_absolute_error}) and importance weights $r_s^{\method, \method^*}$ (Eq.~\ref{eq: computable_is_ratios}) using approximation method $\method^*$. Fit $\hat{k}$ as explained in Section~\ref{s: psis}.
  \item Increase the accuracy of $\method^*$ and repeat Step 3 until $\text{MAE}^{\method, \method^*}$ and $\hat{k}$ converge. If $\hat{k}$  converges to a value larger than $0.7$, increase the accuracy of $\method$ and go back to Step 2.
  \item Compute any posterior estimates using Eq.~\ref{eq: importance_sampling}, with $r_s$ being the values to which $r_s^{\method, \method^*}$ finally converged.
\end{enumerate}
The algorithm can be used in two ways, depending on how Step 1 is done. Firstly, it can be used to validate the reliability, and correct the errors of a given method $\method$, which can be for example a software default. Secondly, a smart initial selection of $\method$ can provide substantial speed gains compared to often rather conservatively set software defaults, while still maintaining reliability of the inferences.

In the latter case, we generally recommend selecting $\method$ initially so that sampling is as fast as possible. For example, for non-adaptive ODE solvers $\method = \method(h)$, one can first try using the largest sensible step size $h$ that does not result in immediate failure. Sometimes the solver returns infinite or NaN values or values which are inconsistent with the observation model\footnote{For example negative values when positive ones are required.} and MCMC cannot proceed, but in this case the sampler will fail fast and not much time is wasted. Selecting a good $\method$ is more difficult in the case of adaptive solvers $\method = \method(\tol)$ whose adaptation rules are discontinuous and whose gradients also need to be approximated. In their case, using too large tolerances $\tol$ cause inaccurate gradients and ragged likelihood surfaces, which can cause MCMC to struggle and become slower, and these issues are slower to detect. We discuss this effect in more detail in Appendix~\ref{s: impact_on_sampling_eff}.

\section{Results}
We developed an R-package called \textbf{odemodeling} for fitting Bayesian ODE models using Stan \citep{StanTeam} and it is available at \url{https://github.com/jtimonen/odemodeling}. In addition to the adaptive built-in ODE solvers of Stan, it implements two explicit Runge-Kutta methods using a fixed but tuneable number of steps. Furthermore, it implements our workflow for determining reliability of ODE model inference and other convenience functions.

We demonstrate our proposed workflow using various ODE solvers and two ODE models, first with simulated and then with real data. In all experiments we use the dynamic HMC algorithm \citep{betancourt2018,StanTeam} implemented in Stan for MCMC sampling. We use initial leap frog\footnote{This refers to the step size of the MCMC sampler, and should not be confused with the step size of the ODE solver used to solve the modeled system. The lower initial value of 0.1 was observed to reduce some warmup problems compared to default 1.0.} step size $0.1$ and otherwise default configuration, unless otherwise mentioned. In each experiment we run 4 independent MCMC chains for 4000 iterations and the first 2000 iterations are discarded as warmup. Code for reproducing the experiments is available at \url{https://github.com/jtimonen/numapprox-is}. Full experiment details are in Appendix~\ref{s: experiment_setups}.

\begin{figure}[!h]
    \centering
    \includegraphics[width=0.9\textwidth]{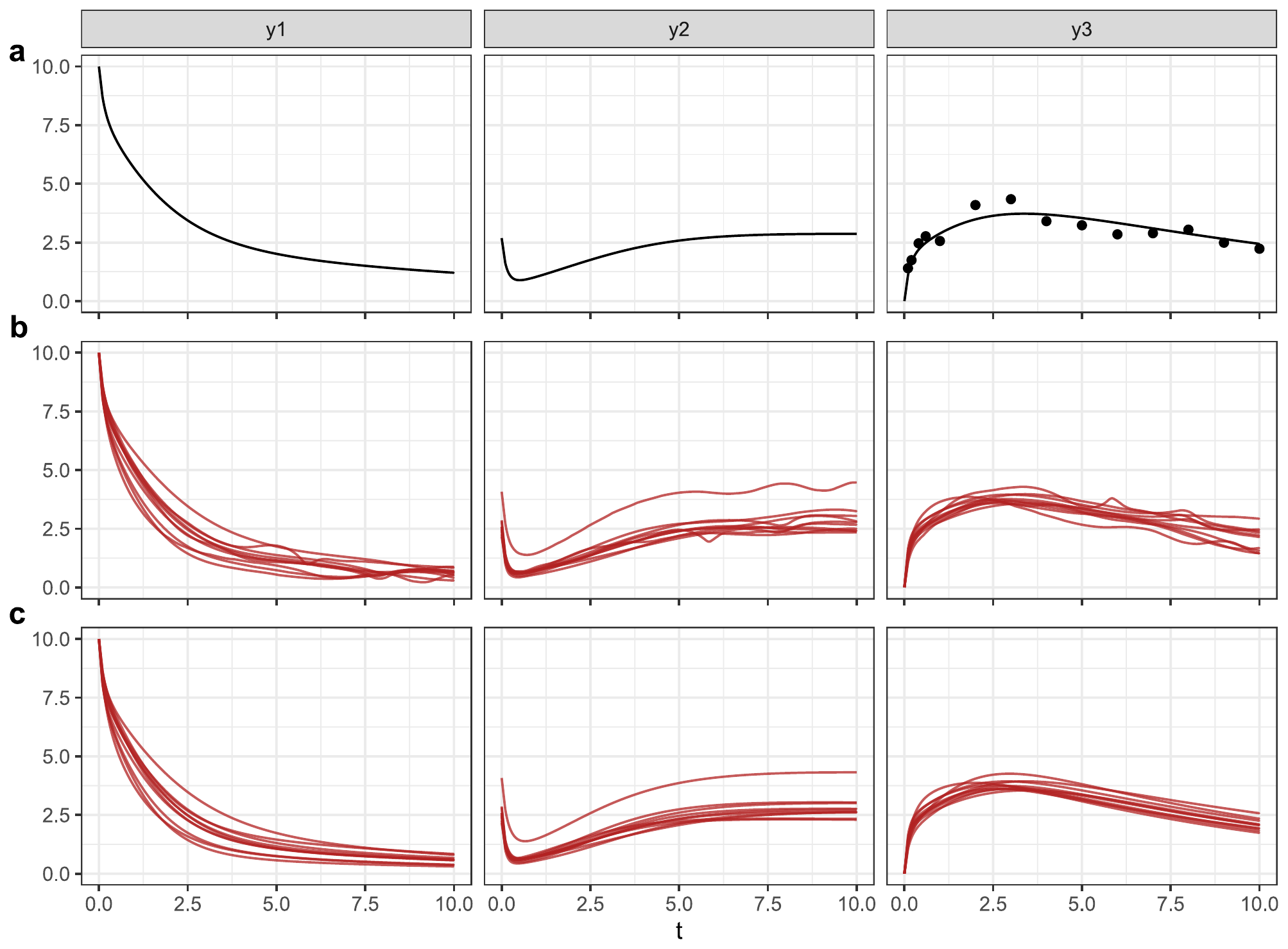}
    \caption{Illustration of the ODE system used in the TMDD experiment. The three columns correspond to (concentrations of) the ligand $y_1$, receptor $y_2$, and receptor-ligand complex $y_3$. \textbf{a)} The black line shows the ODE trajectory used in data simulation, solved using $\text{BDF}(10^{-15})$. The black dots are the generated noisy data. 
    \textbf{b)} ODE solutions solved using $\text{BDF}(0.02)$, corresponding to 10 parameter draws from the posterior induced by $\text{BDF}(0.02)$. \textbf{c)} ODE solutions corresponding to the same 10 posterior parameter draws as in \textbf{b)}, but solved using $\text{BDF}(10^{-12})$. The black and red lines visualized the ODE solutions at a dense set of time points $t=0.00, 0.10, 0.11, 0.12, \ldots, 10.00$. See detailed explanation of the BDF method and the used implementation in Appendix~\ref{s: bdf_details}.}
    \label{fig: tmdd_simulation}
\end{figure}

\subsection{Target-mediated drug disposition model}
In the first experiment, the model is a target mediated drug disposition (TMDD) model \citep{mager2001, aston2011}. This is a common pharmacokinetic-pharmacodynamic model for reversible binding of drug/ligand ($y_1$) with receptor ($y_2$), where a receptor-ligand complex ($y_3$) is formed. For solving the ODE system, we use the backward differentiation formulae method (BDF, see Appendix~\ref{s: ode_solver_details}). We denote the BDF solver with absolute and relative tolerances $\atol = \rtol = \tol$ by BDF($\tol$). The ODE solution $\odevar^{\text{BDF}(10^{-15})}(t)$ using the simulation parameters (Appendix~\ref{s: experiment_setups}) and the simulated noisy data used in this experiment are visualized in Fig.~\ref{fig: tmdd_simulation}a.

\begin{figure}[!h]
    \centering
    \includegraphics[width=\textwidth]{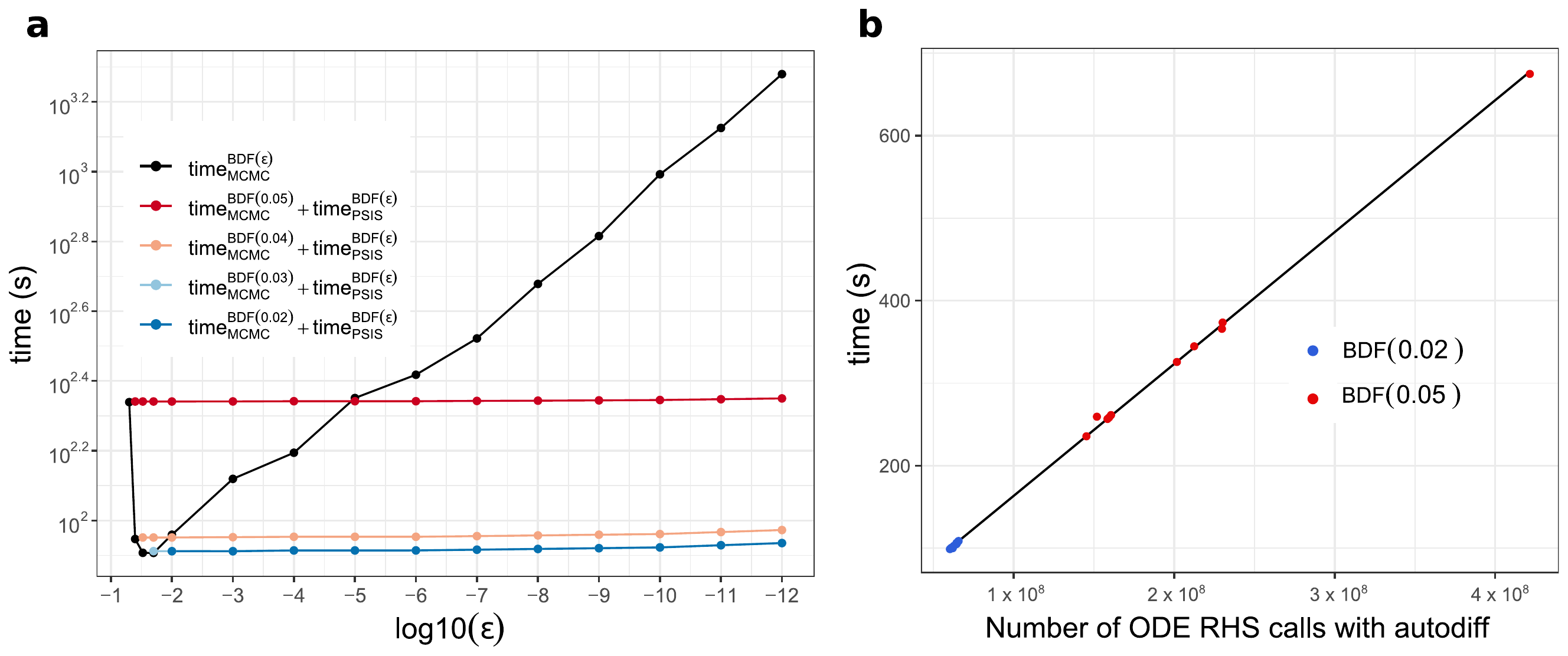}
    \caption{Runtime comparison in the TMDD experiment. \textbf{a)} The black line corresponds to running MCMC directly with $\method=\text{BDF}(\tol)$, where $\tol$ is on the $x$-axis. The four colored lines correspond to first running MCMC with $\method= \text{BDF}(0.05)$, $\text{BDF}(0.04)$, $\text{BDF}(0.03)$ or $\text{BDF}(0.02)$, and additionally computing the importance weights needed to correct resulting posterior estimates as if the draws were from $\method=\text{BDF}(\tol)$, where $\tol$ is on the $x$-axis. The same MCMC algorithm and same number of chains and draws is used in all MCMC sampling.
    \textbf{b)} The MCMC runtime is almost completely explained by the number of times we need to perform automatic differentiation for the ODE RHS function. The red and blue dots each correspond to one MCMC chain, x-axis showing the number of needed RHS calls with AD and y-axis the chain runtime.}
    \label{fig: tmdd_timing}
\end{figure}

\begin{figure}[!h]
    \centering
    \includegraphics[width=\textwidth]{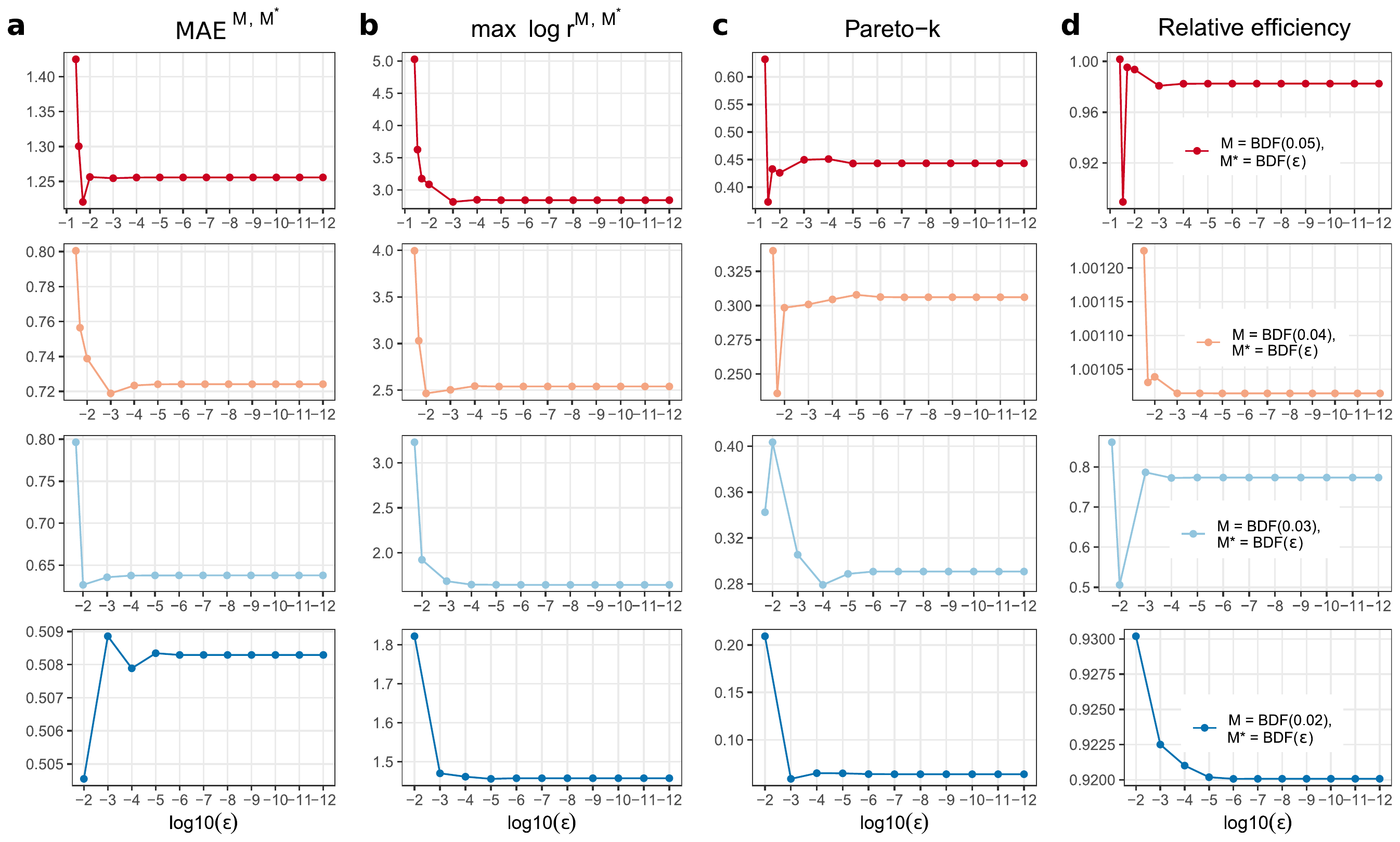}
    \caption{Convergence of different metrics as $\method^*$ is made more and more accurate in the TMDD experiment.  The \textbf{a)} column shows the maximum absolute error (MAE) between posterior ODE solutions $\odevar^{\method}$ and $\odevar^{\method^*}$ (Eq.~\ref{eq: max_absolute_error}), as a function of the tolerance of $\method^*$. The \textbf{b)} column shows the largest importance ratio $r^{\method, \method^*}$.
     The \textbf{c)} column shows for Pareto smoothed importance sampling the $\hat{k}$ diagnostic, and \textbf{d)} column its estimated relative efficiency. The rows correspond to different choices of the method $\method$ (see the legend in column \textbf{d}) used during MCMC.}
    \label{fig: tmdd_metrics}
\end{figure}

To study the effect of tolerances, we first run MCMC sampling using $\method = \text{BDF}(\tol)$ different $\tol$ values. Example ODE solutions from the posterior using $\tol=0.02$ are shown in Fig.~\ref{fig: tmdd_simulation}b, and corresponding solutions with $\tol=10^{-12}$ in Fig.~\ref{fig: tmdd_simulation}c. The tolerance value $\tol=0.02$ is very high and the solutions look unstable compared to $\tol=10^{-12}$.

The timing results in Fig.~\ref{fig: tmdd_timing} (black line) show that for sufficiently small values (roughly $\tol < 0.02$) of $\tol$, the MCMC runtime increases consistently as $\tol$ is decreased. For example, with $\tol = 10^{-10}$, which is the default in Stan, MCMC sampling takes around $10^3$ seconds. On the other hand, we can sample from the same distribution by first MCMC sampling with $\method = \text{BDF}(0.02)$ and then performing importance sampling with $\method^* = \text{BDF}(10^{-10})$ in just $10^2$ seconds. This time comes almost solely from the MCMC sampling, and we could change $\method^*$ to even more accurate, with virtually no extra cost. We note that plain runtime of MCMC is not always a good measure of sampling efficiency. In this case, however, it is informative as we always use the same MCMC algorithm, the bulk and tail estimated effective sample sizes \citep{vehtari2021rank} are very similar in all cases, and convergence diagnostics are satisfactory (see Table~\ref{t: mcmc_diagnostics_tmdd}).

Fig.~\ref{fig: tmdd_simulation}b-c demonstrated that the $\text{BDF}(0.02)$ gives suspicious ODE solutions, indicating that we cannot trust it as such and need our reliability workflow. However, quantities in Fig.~\ref{fig: tmdd_metrics} validate that importance sampling can be trusted for all tested tolerances in $\method$, because the estimate of the shape parameter of the generalized Pareto distribution $\hat{k}$ stabilizes at a value smaller than $0.5$ as $\method^*$ is made more and more accurate. Performing this reliability check takes an insignificantly small amount of time compared to MCMC sampling. The estimated relative efficiency of the importance sampling phase is close to 1 (Fig.~\ref{fig: tmdd_metrics}d), and we do not significantly lose efficiency due to it.

\begin{figure}[!h]
    \centering
    \includegraphics[width=\textwidth]{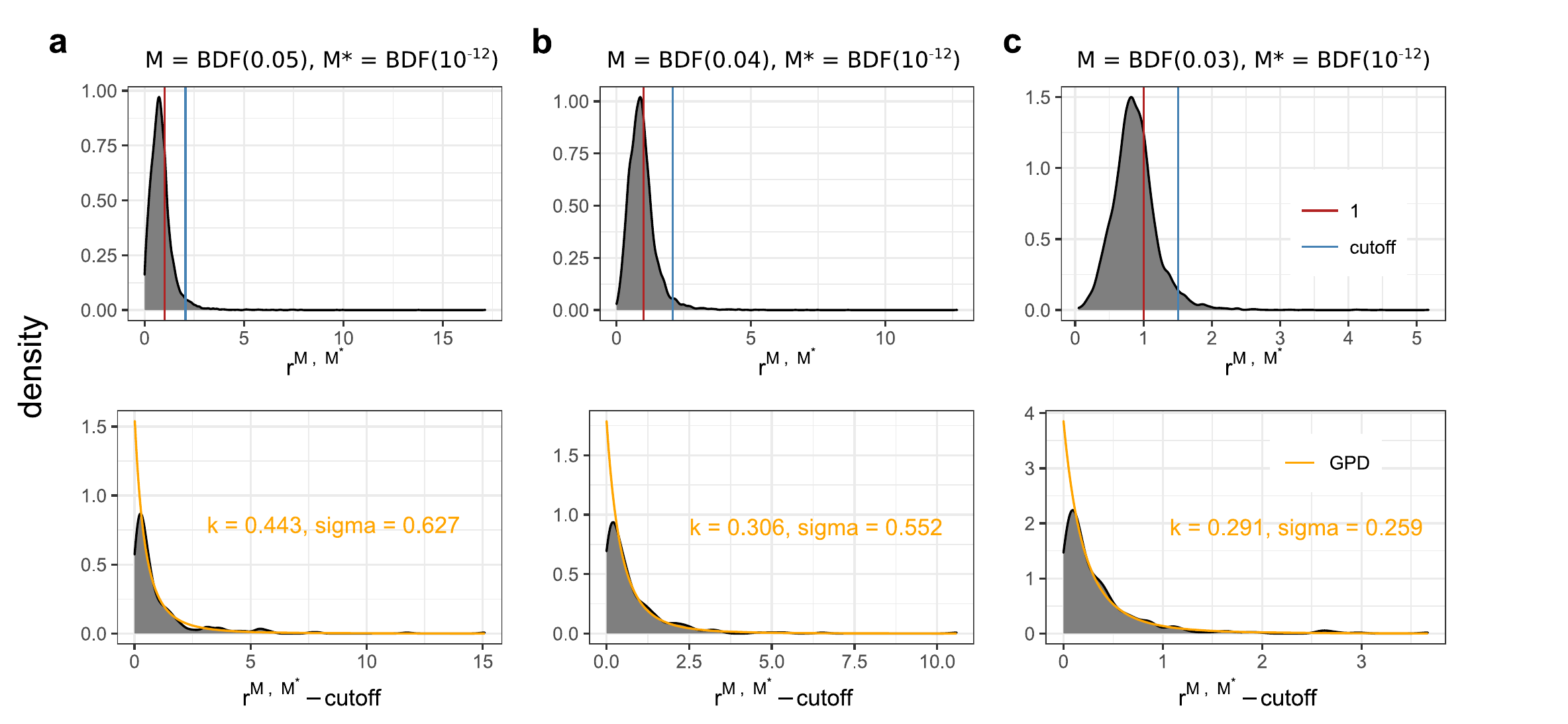}
    \caption{Illustration of fitting the generalized Pareto distribution in the TMDD experiment with $\method^* = \text{BDF}(10^{-12})$. \textbf{a)} $\method = \text{BDF}(0.05)$. \textbf{b)} $\method = \text{BDF}(0.04)$. \textbf{c)} $\method = \text{BDF}(0.03)$.}
    \label{fig: tmdd_psis}
\end{figure}

Fig.~\ref{fig: tmdd_psis} illustrates the distribution of the importance ratios and fitting the generalized Pareto distribution (GPD) in different cases. We see that the tail of the ratios becomes less and less thick as tolerances of $\method$ are decreased.

The runtime starts increasing if the tolerances are made too large (approx. $\tol > 0.03)$. This means that we cannot start our workflow with $\tol$ being arbitrarily large. To understand the reason for this, we have recorded some HMC NUTS metrics in Table~\ref{t: mcmc_diagnostics_tmdd}. The runtime is explained almost completely by the number of ODE RHS evaluations with AD (Fig.~\ref{fig: tmdd_timing}b), and too large tolerances cause the leap frog step size to adapt to smaller values, meaning that more leap frog steps and therefore more RHS evaluations with AD are needed. Potential reasons for this can be that the discontinuity of the ODE solver step size adaptation rule becomes more evident with large tolerances, which makes the likelihood surface more discontinuous and complicates sampling. The problems are amplified by the fact that the gradient error grows when tolerances are increased (See Section~\ref{s: ode_sens} and  Appendix~\ref{s: impact_on_sampling_eff}).

\subsection{Lotka-Volterra model}
In the second experiment, we study a model of predator-prey dynamics between Canadian lynx and snowshoe hare. See Appendix~\ref{s: experiment_setups} for details about the model and data. In this experiment, we first use the adaptive RK45 ODE solver (Appendix~\ref{s: ode_solver_details}), and use $\text{RK45}(\tol)$ to denote it with tolerances $\rtol = \atol = \tol$.

\begin{figure}[!h]
    \centering
    \includegraphics[width=\textwidth]{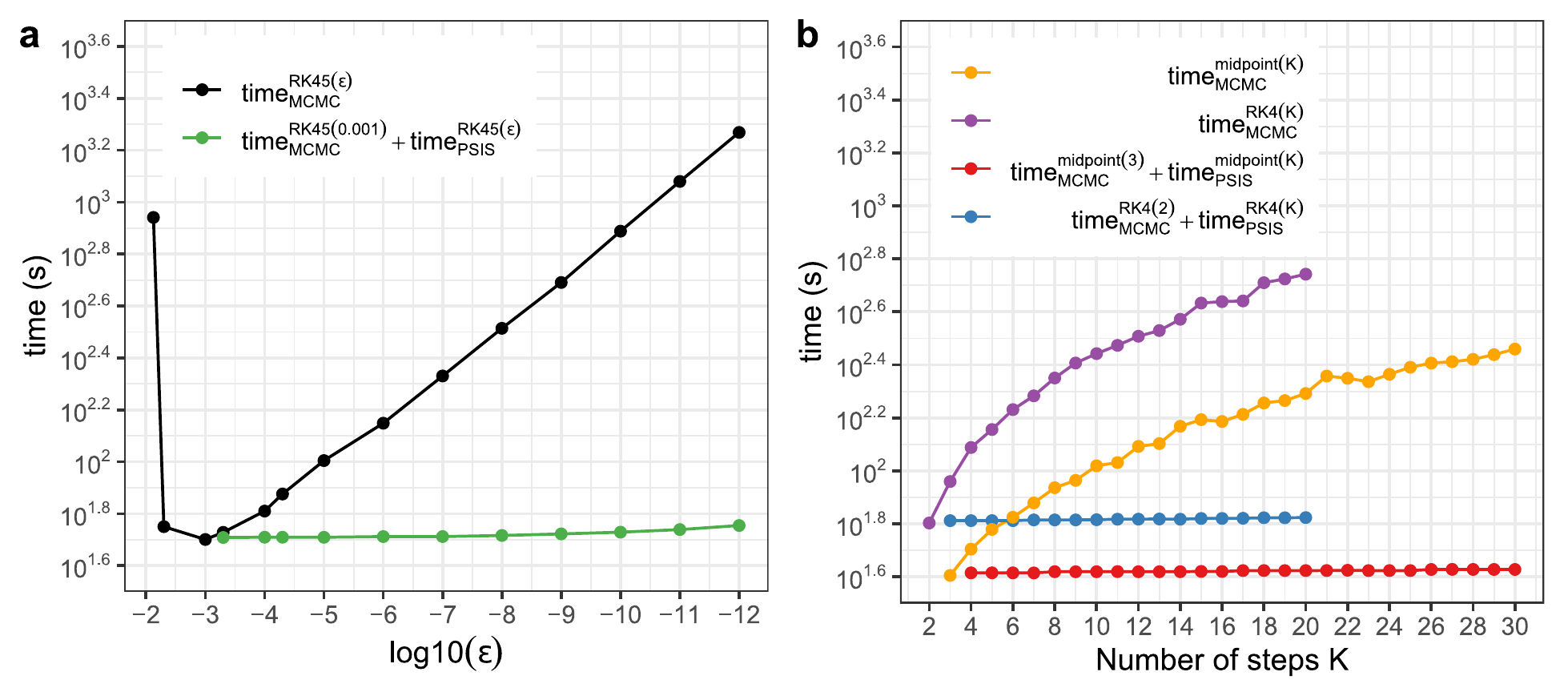}
    \caption{Runtime results in the Lotka-Volterra experiment. \textbf{a)} The black line corresponds to just running MCMC directly with $\method=\text{RK45}(\tol)$, where $\tol$ is on the $x$-axis. The green line corresponds to first using  MCMC with $\method=\text{RK45}(0.001)$, and additionally computing the importance weights against the posterior density induced by $\text{RK45}(\tol)$.
    \textbf{b)} Using the explicit midpoint and RK4 methods, with fixed number of steps $K$ between output time points.}
    \label{fig: lv_timing}
\end{figure}

\begin{figure}[!h]
    \centering
    \includegraphics[width=\textwidth]{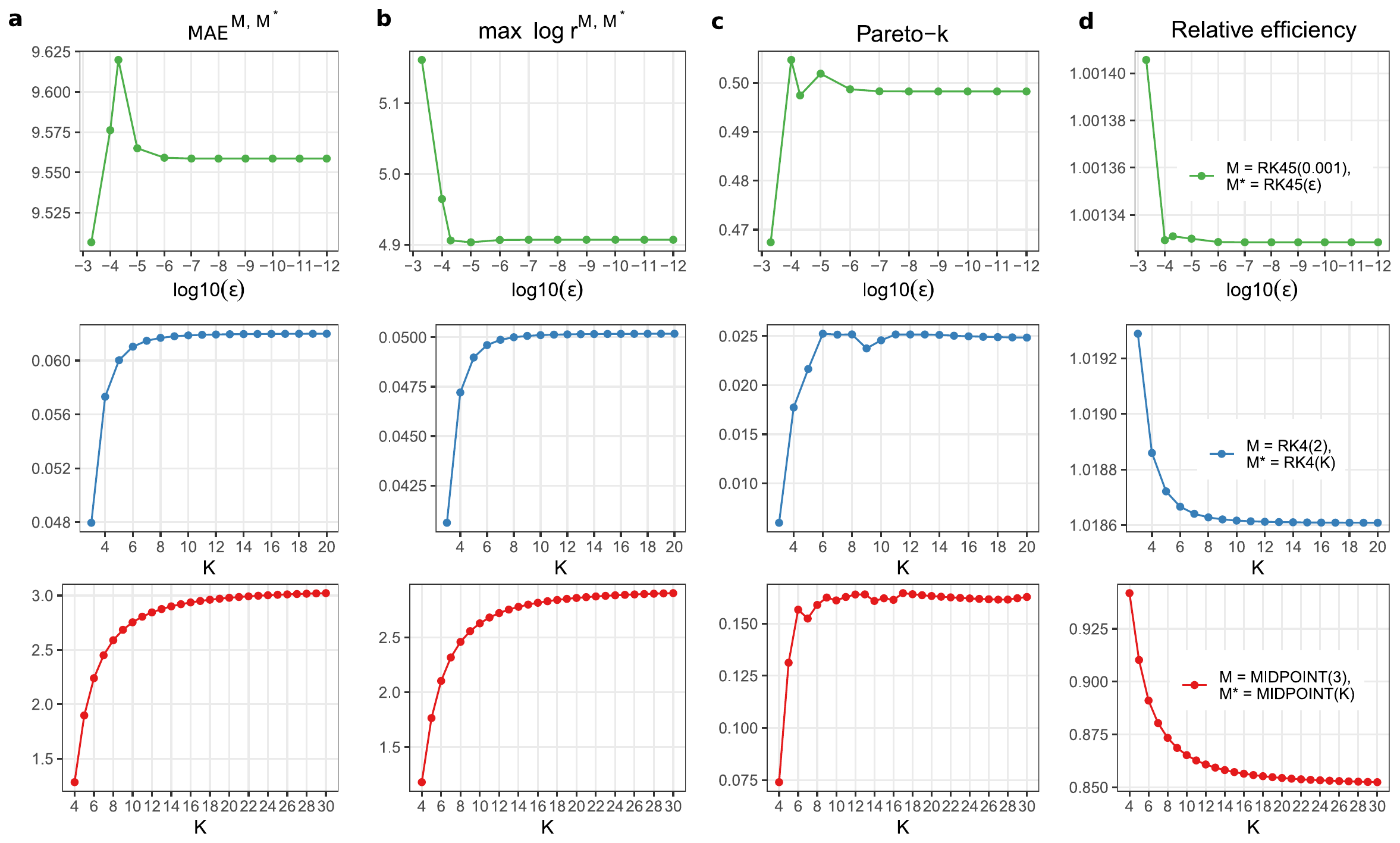}
    \caption{Convergence of different metrics during our workflow in the Lotka-Volterra experiment, as the method $\method^*$ is made more and more accurate. The rows correspond to different choices of the initial ODE solver $\method$ (see the legend in column \textbf{d}) used during MCMC.}
    \label{fig: lv_metrics}
\end{figure}

 We first run MCMC sampling using $\method = \text{RK45}(\tol)$ with different $\tol$ values. The results in Fig.~\ref{fig: lv_timing}a (black line) show that for small values (approx $\tol < 10^{-3}$) of $\tol$, the runtime increases as $\tol$ is decreased. For example, with $\tol = 10^{-6}$, which is the default in Stan, MCMC sampling takes around $10^{2.15} \approx 140$ seconds. On the other hand, we can sample from the same distribution by first MCMC sampling with $\method = \text{RK45}(0.001)$ and then performing importance sampling with $\method^* = \text{RK45}(10^{-6})$ in just $10^{1.7} \approx 50$ seconds.

We performed MCMC inference and importance sampling using also the RK4 and midpoint methods for solving the ODE (Appendix~\ref{s: ode_solver_details}). We use notation $\text{RK4}(K)$ and $\text{midpoint}(K)$ where $K$ is the number of steps taken between subsequent output time points and thus implicitly determines the step size $h$. The results in Fig.~\ref{fig: lv_timing}b (purple and orange lines) show that for both methods, the runtime increases consistently with $K$. Furthermore, we find that using $\method = \text{midpoint}(3)$ is faster than any tested tolerance values for RK45 and $\method = \text{RK4}(2)$ is a bit slower than RK45 with $\tol = 0.001$ or $\tol = 10^{-4}$. Smaller $K$ for either midpoint or RK4 in this case caused the solver to break at initial parameter values and not be able to produce MCMC draws. 

 Quantities in Fig.~\ref{fig: lv_metrics} validate that our importance sampling workflow can be trusted. The $\text{MAE}^{\method, \method^*}$ and other metrics converge as $\method^*$ is made more accurate, and $\hat{k}$ values converge to $< 0.5$ in all cases with close to 1 relative efficiency. The convergence is especially smooth for the non-adaptive solvers, as $K$ is increased for $\method^*$.

\section{Discussion}
We have shown that the proposed approach is useful for efficient and reliable Bayesian inference of general ODE models. We have demonstrated our workflow in ODE problems, but it is generally applicable also to various other types of models that require approximate numerical methods. 

We have shown that by using a less accurate numerical method $\method$ during MCMC and importance sampling with increasingly more accurate versions $\method^*$ of the method can be an order of magnitude faster than plain MCMC sampling with a software default method $\method^*$. Moreover, after MCMC sampling with some method $\method$, users can easily and without further computational effort check what the posterior of ODE solutions would look like with any given other solver.

Efficiently selecting the initial method $\method$ to be used in our workflow still has some challenges if $\method$ has to belong to the class of commonly used adaptive ODE solvers. Inference with these methods can become slower as $\method$ is made too inaccurate. These methods have not originally been designed to be used in Bayesian inference, but have been widely adopted into even gradient-based MCMC and optimization software despite their discontinuous adaptation rules, and the fact that their local error control itself is not enough for reliable statistical inference.

 Our framework enables also rapid testing of different types of new and old numerical solvers, which do not need to have error control. The only requirement is that they are convergent numerical methods whose accuracy can be controlled. We have shown a large potential and obtained promising results using explicit non-adaptive solvers whose gradient is computed using direct automatic differentiation of the solver formula.

\subsection*{Acknowledgements}

We would like to acknowledge the computational resources provided by Aalto Science-IT, Finland. This work was supported by the Academy of Finland Flagship programme: Finnish Center for Artificial Intelligence, and the Academy of Finland projects 340721, 311584, and 328401.

\clearpage
\bibliography{references}

\clearpage
\appendix

\vspace{1.2cm}


\section{ODE solvers}
\label{s: ode_solver_details}
Here we give the details about the used ODE solvers and their used implementations. We implemented the explicit non-adaptive RK methods as user-defined functions in Stan. The RK45 and BDF solvers are built into Stan, and we describe how Stan uses the Boost Odeint \citep{ahnert2011} and \textsc{SUNDIALS} \citep{hindmarsh2005} libraries to obtain both the ODE solutions and their sensitivities. The goal is to focus on how the specified tolerances and other factors affect the approximate solution and sensitivities. We used Stan version 2.28, which depends on Boost version 1.75.0 and \textsc{SUNDIALS} version 5.7.0.

\subsection{Explicit Runge-Kutta methods}
\label{s: rk4_midpoint_details}
A general $R$-stage explicit RK method with step size $h$ uses the update rule
\begin{equation}
    \label{eq: explicit_rk}
    \odevar^{\text{RK}}(\tau_{j+1}) = \odevar^{\text{RK}}(\tau_{j}) + h \sum_{r=1}^R b_r \bm{k}_r, \hspace{1cm} j \geq 1,
\end{equation}
where
\begin{align*}
    \bm{k}_1 &= f_{\psi}\left(\odevar^{\text{RK}}(\tau_{j}), \tau_{j} \right) \\
    \bm{k}_2 &= f_{\psi}\left(\odevar^{\text{RK}}(\tau_{j}) + h \cdot a_{2,1} \bm{k}_1, \tau_{j} + c_2 h \right) \\
    \bm{k}_3 &= f_{\psi}\left(\odevar^{\text{RK}}(\tau_{j}) + h \cdot (a_{3,1} \bm{k}_1 + a_{3,2} \bm{k}_2), \tau_{j} + c_3 h \right) \\
    \vdots \\
     \bm{k}_R &= f_{\psi}\left(\odevar^{\text{RK}}(\tau_{j}) + h \cdot (a_{R,1} \bm{k}_1 + \ldots + a_{R,R-1} \bm{k}_{R-1}), \tau_{j} + c_{R} h \right),
\end{align*}
and $a_{r,l}$, $b_r$, and $c_r$ are suitable coefficients. For the MIDPOINT method,  the number of stages is $R = 2$ and the coefficients are $a_{21} = \frac{1}{2}$, $b_1 = 0$, $b_2 = 1$ and  $c_2 = \frac{1}{2}$. For RK4, the number of stages is $R=4$ and the coefficients  are $a_{21} = \frac{1}{2}$, $a_{31} = 0$, $a_{32} = \frac{1}{2}$, $a_{41} = 0$, $a_{42} = 0$, $a_{43} = 1$, $b_1 = \frac{1}{6}$, $b_2 = \frac{1}{3}$, $b_3 = \frac{1}{3}$, $b_4 = \frac{1}{6}$ and $c_2 = \frac{1}{2}$, $c_3 = \frac{1}{2}$, $c_4 = 1$. 

\subsection{RK45}
\label{s: rk45_details}
Methods with adaptive step size control can have a varying step size $h_j$. An example is the RK45 method \citep{dormand1980}, which estimates its local truncation error and adapts $h_j$ so that a certain requirement based on given absolute ($\atol$) and relative ($\rtol$) tolerances is satisfied. On each step $j$, a test solution for the next state is computed using two RK formulas, which have order (number of stages) $4$ and $5$. We denote the current approximation at $t_{j}$ by $\odevar^{\text{RK45}}_j \in \mathbb{R}^D$ and the test solution given by order $R$ method at the next point $t_{j+1} = t_{j} + h_j$ by $\odevar_{j+1}^{(R)}$. 

Stan uses the implementation in Boost odeint\footnote{\url{https://www.boost.org/doc/libs/1_75_0/libs/numeric/odeint/doc/html/}}, where the step size control is based on the maximum relative estimated local truncation error
\begin{equation}
\label{eq: boost_normed_error}
    \normedError_j = \max_{d=1, \ldots, D} \left \{ \frac{ |y_{j+1,d}^{(5)} - y_{j+1,d}^{(4)}| } { \atol + \rtol \cdot \left( |y^{\text{RK45}}_{j,d}| + h_j \cdot |f_{j,d}|\right) ) } \right \} ,
\end{equation}
where $f_{j,d}$ is the $d$th component of $f(\odevar^{\text{RK45}}_j, t_j)$. If $\normedError_j > 1$, the test step is rejected as the step size has to be decreased to satisfy the tolerances. This is done by setting
\begin{equation}
\label{eq: boost_rk45_stepsize_decrease}
     h_j \leftarrow h_{j} \cdot \max \left\{ 0.9 \cdot \normedError_j^{-1 / 3}, 1/5 \right\}
\end{equation}
recomputing $v_j$ until $v_j \leq 1$. Once an acceptable step size $h_j$ is found, the higher order test solution is then used to advance the integrator, i.e. $\odevar^{\text{RK45}}_{j+1} = \odevar_{j+1}^{(R+1)}$. The next step size is then
\begin{equation}
\label{eq: boost_rk45_stepsize_update}
 h_{j+1} =
\begin{cases}
    h_{j} \cdot \min \left \{ 0.9 \cdot \normedError_j^{-1 / 5}, 5 \right \} &, \normedError_j < 0.5\\ 
    h_{j} &,  0.5 \leq \normedError_j \leq 1,
 \end{cases}
\end{equation}
where the first case means that the step size increases. These update rules are based on theory about the optimal step size updating, and safety modifications that ensure that step size does not change too fast (cf. \cite{hairer1993}). There exist automatic ways for selecting a good initial step size $h_{0}$ \citep{hairer1993}, but Stan does not attempt that and always sets $h_{0} = 0.1$.

Stan uses the so called dense output version of RK45, for which the maximum step size is not restricted based on the time points where the solution is requested, until for the very last output time point. Instead, an interpolation method \citep{dormand1986} is used to obtain the solution at the intermediate output time points. However, there is a limit for maximum number of steps that can be taken between two consecutive output time points, for which we used the value $10^5$. If this limit is reached during MCMC, the current proposal is rejected.

Continuous forward sensitivity analysis of the RK45 solutions is in Stan implemented by using the above strategy as such for the extended system that involves also the forward sensitivity equations (see Section~\ref{s: ode_sens}). This means that also the extended system dimensions affect the error control and step size updates similarly as the original dimensions.

\subsection{BDF}
\label{s: bdf_details}
The backward differentiation formulae (BDF) method is a linear multistep method, and Stan uses the implementation in the \textsc{CVODES} package  \citep{hindmarsh2021} of the \textsc{SUNDIALS} software suite. Each step $j$ is based on the rule
\begin{equation}
\odevar^\text{BDF}_{j} = \sum_{i = 1}^{q} a_{j,i} \odevar^\text{BDF}_{j-q} + h_j \beta_{j} f(\odevar^\text{BDF}_j, t_j),
\end{equation}
where $\odevar^\text{BDF}_j = \odevar^\text{BDF}(t_{j})$. The order $q$ can vary between 1 and 5, and is adapted simultaneously with the step size $h_j$. The coefficients $\alpha_{j,i}$, $\beta_{j}$ are determined by the recent history of the step sizes \citep{byrne1975, jackson1980}. For this method, estimation of the local truncation error and adaptation rules are highly complex and all details are described in \citep{hindmarsh2021}. A nonlinear rootfinding problem must be solved at each step to obtain $\odevar^\text{BDF}_j$, and this is done approximately using modified Newton iteration. This in turn requires solving a linear system, for which Stan uses the dense linear solver facilities of \textsc{SUNDIALS}. In addition to controlling the estimated local truncation error, also the Newton iteration convergence has to be controlled. The stopping criterion for the Newton iterations is in \textsc{CVODES} attempted to set so that its error does not interfere with local error control.

The usage of BDF in Stan involves a hard coded initial step size setting rule which depends on the first requested output time. One effect of this is that the returned solution at a given time point depends not only on the selected tolerances, but also on whether the solution output has been requested at earlier time points. This is why when visualizing the ODE solutions at a dense set of time points in Figure~\ref{fig: tmdd_simulation}, we are not able to solve the system at $0 < t < 0.1$, because $t = 0.1$ is the first observation time point. Requesting a dense solution also before the first data time point would cause the ODE solutions at the data time points differ from those we get when we solve only at the data time points (i.e. what we do when fitting the model).


Also this solver has a limit for the maximum number of steps that can be taken between two consecutive output time points, and we used the value $10^5$.

Continuous forward sensitivity analysis of the BDF solutions is in Stan implemented by using the continuous sensitivities provided by \textsc{CVODES}. They are computed so that the same linear multistep formula is used also for solving the sensitivity equations. Stan selects to employ the staggered corrector method \citep{feehery1997}, where a separate Newton iteration is used to solve the sensitivity system after the convergence of Newton iteration for the original system. The sensitivity dimensions are included in the error control, which means that they affect the adaptation. The relative tolerance for sensitivity dimensions is the same $\rtol$ as for the original dimensions, but their absolute tolerances are automatically scaled versions of $\atol$. See more details in \cite{hindmarsh2021}.

\section{Impact of numerical approximations on sampling efficiency}
\label{s: impact_on_sampling_eff}

Numerical approximation procedures may introduce issues for gradient based MCMC methods such as HMC and its variants. These methods aim to reduce
the autocorrelation of subsequent draws by numerically simulating the trajectory
of a fictitious particle which is accelerated by the negative gradient of the unnormalized log posterior density. The sampling efficiency  depends on the average energy error of the simulated particle and
decreases sharply if the numerical simulation of its trajectory is not accurate enough \citep{neal2011}. However, it is usually assumed that the average energy error can be brought arbitrarily close to zero by lowering the step size used for the numerical simulation of the fictitious particle, and this assumptions is usually relied upon to optimize sampling efficiency \citep{StanTeam}.

Numerical procedures can cause this assumption to fail in two ways. If the procedure has an adaptive control flow (e.g. iterate until some value passes some threshold), the numerical procedure may introduce \emph{discontinuities} in the posterior density. 
If the numerical procedure includes continuous adaptivity (e.g. continuously increasing or decreasing the step size) or if the computation of the sensitivities of its approximate result uses a mathematical identity which only holds for the exact result (e.g. when using the implicit function theorem), then the numerical procedure may introduce a \emph{gradient mismatch}, meaning that the vector field which gets used to accelerate the fictitious particle is not equal to the gradient of the (approximate) log posterior density.

While the direct method generally does not introduce a gradient mismatch, it can still introduce discontinuities if it is applied to an adaptive method. Forward continuous sensitivity analysis generally introduces both a gradient mismatch and discontinuities if applied to adaptive ODE solvers. However, if it is applied to an explicit ODE solver with a fixed step size configuration it need not do either. Except for edge cases or if handled in a special way, any implicit ODE solver generally introduces both a gradient mismatch and discontinuities, as does continuous adjoint sensitivity analysis.

Whether gradient mismatches or discontinuities cause gradient based MCMC methods to struggle depends mainly on the size of the two effects. HMC and its variants can generally tolerate minor discontinuities or gradient mismatches, if the average energy error can still be made sufficiently small. Generally, stationarity of the (approximate) target distribution is not lost due to discontinuities or gradient mismatches, as neither influences the reversibility or the phase space volume preservation of the commonly used leapfrog integrator \citep{hairer2003geometric}, which are the main necessities for stationarity of the target distribution \citep{neal2011}.

\section{Experiment setups}
\label{s: experiment_setups}
\subsection{TMDD model}
The dynamics are modeled using the three-dimensional ODE system
\begin{equation}
f_{\psi}(\odevar, t) = 
    \begin{bmatrix}
    - k_{\text{eL}} y_1 - (k_{\text{on}} y_1 y_2 - k_{\text{off}} y_3) \\
    k_{\text{in}} - k_{\text{out}} y_2 - (k_{\text{on}} y_1 y_2 - k_{\text{off}} y_3) \\
    (k_{\text{on}} y_1 y_2 - k_{\text{off}} y_3) - k_{\text{eP}} y_3
    \end{bmatrix}
\end{equation}
which has six unknown parameters $\odeparam = \left\{k_{\text{on}}, k_{\text{off}}, k_{\text{in}},  k_{\text{out}}, k_{\text{eL}}, k_{\text{eP}} \right\}$. The initial state is $\odevar^{(0)} = \odevar(t_0) = \left[L_0, \frac{k_{\text{in}}}{k_{\text{out}}}, 0\right]^\top$, where $t_0 = 0$ and $L_0 = 10$ is the initial drug bolus.

The data $\dat = \{x^{(n)} \}_{n=1}^{15}$ consist of measurements $x^{(n)}$, which are the measured amount of complex $y_3$ at time points $t_n =$ 0.1, 0.2, 0.4, 0.6, 1, 2, 3, 4, 5, 6, 7, 8, 9, 10. The likelihood (sampling distribution) is
\begin{equation}
    \label{eq: likelihood_tmdd}
    p\left(\dat \mid \param \right) = \prod_{n=1}^N  \text{Normal}\left(x^{(n)} \mid y_3(t_n), \sigma^2 \right),
\end{equation}
where the noise magnitude $\sigma > 0$ is an unknown parameter. All model parameters are therefore $\param =  \left(\odeparam, \sigma \right).$ They are given priors
\begin{align*}
    k_{\text{on}}, k_{\text{eL}} &\sim \text{LogNormal}(-1, 0.3) \\
    k_{\text{off}}, k_{\text{in}}, k_{\text{out}} &\sim \text{LogNormal}(0, 0.3) \\
    k_{\text{eP}} &\sim \text{LogNormal}(-3, 0.3) \\
    \sigma &\sim \text{LogNormal}(0, 0.3).
\end{align*}
We generate artificial data $\dat$ from the sampling distribution specified by Eq.~\ref{eq: likelihood_tmdd}, using parameter values $\odeparam = \left\{0.592, 0.900, 2.212, 0.823, 0.201, 0.024\right\}$ \citep{aston2011} and  $\sigma = 0.5$. In data simulation, the ODE system is solved using the BDF solver with $\atol = \rtol = 10^{-15}$. When fitting the model, the MCMC chains are initialized at point $\param = \bm{1}$.

\subsection{Lotka-Volterra model}
We use the same data and model as \cite{carpenter2018}. The data $\dat = \{\bm{x}^{(n)} \}_{n=0}^{20}$ consists of measurements $\bm{x}^{(n)} = [x_1^{(n)}, x_2^{(n)}]^{\top}$, where $x_1^{(n)}$ and $x_2^{(n)}$ are the number of collected lynx and hare pelts in thousands, respectively, at time point $t_n$. The measurements have been made yearly between 1900 and 1920, so that $t_n = 1900 + n$. The dynamics are modeled using the two-dimensional Lotka-Volterra ODE system
\begin{equation}
f_{\psi}(\odevar, t) = 
    \begin{bmatrix}
    \psi_1 y_1 - \psi_2 y_1 y_2 \\
    \psi_3 y_1 y_2 - \psi_4 y_2,
    \end{bmatrix}
\end{equation}
which has four unknown parameters $\odeparam \in \mathbb{R}^4_+$. The likelihood is
\begin{equation}
    p\left(\dat \mid \param \right) = \prod_{n=0}^N  \text{LogNormal}\left(\bm{y}^{(n)} \mid \log \odevar(t_n), \sigma^2 \bm{I}\right)
\end{equation}
where the initial state $\odevar^{(0)} = \odevar(t_0) \in \mathbb{R}^2$ is also an unknown parameter, in addition to the noise magnitude $\sigma > 0$. All model parameters are therefore $\param = \left(\odeparam, \sigma, \odevar^{(0)} \right)$ and they have priors
\begin{align*}
    \psi_1, \psi_3 &\sim \text{Normal}_+(1, 0.5^2) \\
    \psi_2, \psi_4 &\sim \text{Normal}_+(0.05, 0.05^2) \\
    \sigma &\sim \text{LogNormal}(-1, 1) \\
    y_1^{(0)}, y_2^{(0)} &\sim \text{LogNormal}(\log(10), 1).
\end{align*}
In this experiment we initialize the MCMC chains with values $\psi_1, \psi_3, \sigma = 1$, $\psi_2, \psi_4 = 0.1$, and $\bm{y}^{(0)}$ equal to the observation at first time point.

\section{Sampling diagnostics in the TMDD experiment}
MCMC convergence diagnostics \citep{vehtari2021rank} in the TMDD experiments are in Table~\ref{t: mcmc_diagnostics_tmdd}. Corresponding HMC NUTS diagnostics \citep{StanTeam} are in Table~\ref{t: hmc_diagnostics_tmdd}. 

\begin{table}[ht]
\centering
\begin{tabular}{rrrrrr}
  \hline
 & $\tol$ & runtime (s) & max\_rhat & min\_ess\_bulk & min\_ess\_tail \\ 
  \hline
 & 0.05 & 218.4 & 1.002 & 3179.2 & 4309.1 \\ 
   & 0.04 & 88.5 & 1.001 & 3267.8 & 4651.3 \\ 
   & 0.03 & 80.8 & 1.002 & 2643.6 & 4337.1 \\ 
   & 0.02 & 80.7 & 1.001 & 3506.6 & 4609.7 \\ 
   & 0.01 & 91.0 & 1.001 & 2966.9 & 5073.3 \\ 
   & $10^{-3}$ & 131.6 & 1.002 & 3444.4 & 4399.0 \\ 
   & $10^{-4}$ & 156.4 & 1.001 & 3508.4 & 4685.9 \\ 
   & $10^{-5}$ & 224.4 & 1.001 & 3218.6 & 5080.7 \\ 
   & $10^{-6}$ & 261.6 & 1.002 & 3170.9 & 4663.2 \\ 
   & $10^{-7}$ & 332.6 & 1.002 & 3143.8 & 4282.7 \\ 
   & $10^{-8}$ & 476.5 & 1.001 & 3521.8 & 4455.9 \\ 
   & $10^{-9}$ & 653.6 & 1.001 & 3583.0 & 4790.3 \\ 
   & $10^{-10}$ & 982.7 & 1.002 & 3254.0 & 4544.0 \\ 
   & $10^{-11}$ & 1334.5 & 1.001 & 3219.9 & 3788.4 \\ 
   & $10^{-12}$ & 1903.9 & 1.001 & 3420.5 & 4915.2 \\ 
   \hline
\end{tabular}
\caption{MCMC diagnostics in the TMDD experiment. The \texttt{max\_rhat} column refers to the largest $\hat{r}$ value over all sampled parameters, whereas \texttt{min\_ess\_bulk} and \texttt{min\_ess\_tail} refer to the smallest bulk and tail estimated effective sample size, respectively \citep{vehtari2021rank}.} 
\label{t: mcmc_diagnostics_tmdd}
\end{table}

\begin{table}[ht]
\centering
\begin{tabular}{rllllll}
  \hline
 & $\tol$ & accept\_stat & stepsize & treedepth & n\_leapfrog & divergent \\ 
  \hline
 & 0.05 & 0.82 & 0.069 & 5.7 & 70 & 0.00 \\ 
   & 0.04 & 0.87 & 0.14 & 4.6 & 27 & 0.00 \\ 
   & 0.03 & 0.88 & 0.16 & 4.4 & 25 & 0.00 \\ 
   & 0.02 & 0.89 & 0.18 & 4.3 & 23 & 0.00 \\ 
   & 0.01 & 0.92 & 0.17 & 4.3 & 23 & 0.00 \\ 
   & 0.001 & 0.93 & 0.18 & 4.2 & 23 & 0.00 \\ 
   & $10^{-4}$ & 0.93 & 0.19 & 4.2 & 22 & 0.00 \\ 
   & $10^{-5}$ & 0.94 & 0.17 & 4.3 & 24 & 0.00 \\ 
   & $10^{-6}$ & 0.93 & 0.18 & 4.2 & 22 & 0.00 \\ 
   & $10^{-7}$ & 0.92 & 0.19 & 4.1 & 21 & 0.00 \\ 
   & $10^{-8}$ & 0.92 & 0.19 & 4.2 & 21 & 0.00 \\ 
   & $10^{-9}$ & 0.92 & 0.19 & 4.1 & 21 & 0.00 \\ 
   & $10^{-10}$ & 0.94 & 0.17 & 4.3 & 23 & 0.00 \\ 
   & $10^{-11}$ & 0.92 & 0.18 & 4.2 & 22 & $1.3 \cdot 10^{-4}$ \\ 
   & $10^{-12}$ & 0.92 & 0.19 & 4.1 & 21 & 0.00 \\ 
   \hline
\end{tabular}
\caption{HMC NUTS diagnostics in the TMDD experiment. The \texttt{accept\_stat} column is the average the Metropolis acceptance probability over all Hamiltonian trajectories, \texttt{stepsize} is the average leapfrog step size, \texttt{treedepth} is the average depth of the tree used by the NUTS sampler, \texttt{n\_leapfrog} is the average number of leapfrog steps per transition, and \texttt{divergent} is the percentage of divergent transitions out of total post-warmup draws  \citep{StanTeam}.} 
\label{t: hmc_diagnostics_tmdd}
\end{table}


\end{document}